\documentclass[twocolumn,showpacs,preprintnumbers,amsmath,amssymb,nofootinbib]{revtex4-1}
\usepackage{graphicx}
\usepackage{dcolumn}
\usepackage{bm}
\begin{document}
\title{Mesoscopic Description of the Equal Load Sharing Fiber Bundle Model}
\author{Martin Hendrick\email{martin.hendrick@ntnu.no}}
\affiliation{PoreLab, Department of Physics, Norwegian University of Science and Technology, N-7491 Trondheim, Norway}
\author{Srutarshi Pradhan\email{srutarshi.pradhan@ntnu.no}}
\affiliation{PoreLab, Department of Physics, Norwegian University of Science and Technology, N-7491 Trondheim, Norway}
\author{Alex Hansen\email{alex.hansen@ntnu.no}}
\affiliation{PoreLab, Department of Physics, Norwegian University of Science and Technology, N-7491 Trondheim, Norway}
\date{\today {}}
\begin{abstract}
One aim of the equal load sharing fiber bundle model is to describe the critical behavior of failure events. One way of 
accomplishing this, is through a discrete recursive dynamics.  We introduce a continuous mesoscopic equation catching 
the critical behavior found through recursive dynamics. It allows us to link the model with the unifying framework of absorbing phase transitions traditionally used in the study of non-equilibrium phase transitions. Moreover, it highlights the analogy between equal load sharing and spinodal nucleation. Consequently, this work is a first step towards the quest of a field theory for fiber bundle models.
\end{abstract}
\maketitle
\section{Introduction} 
\label{sec:intro_gen}

While equilibrium phase transitions are today well understood, a general framework to study the non-equilibrium counterpart is still
lacking. Recently, major efforts have been invested into identifying the universality classes related to non-equilibrium phase transitions. The theory of absorbing phase transitions (APT) is emerging as a unifying framework (or more generally, field theory applied to non-equilibrium scaling behavior) \cite{hhlp08,t14}. An absorbing phase transition occurs when a system leaves an active state and enters absorbing state from which the system cannot escape by itself.

The APT formalism improved the understanding of the universal behavior of a great variety of models such as epidemic and population
dynamic \cite{dmpt18}, sandpiles \cite{bm08}, interface in random media \cite{lw15}, reaction diffusion systems \cite{t14}. The most prominent and generic non-equilibrium universality class is directed percolation (DP), which is believed to describe phase transition toward an unique abosorbing state of systems that are not characterized by any special symmetry (except, effectively, the rapidity reversal symmetry) or conservation law. This is known as the Janssen-Grassberger directed percolation conjecture \cite{j81,g82}.

Fiber bundle models (FBM) describe rupture phenomena as irreversible fiber breaking processes through discrete breaking rules 
\cite{phc10,hhp15}.  In their simplest form, they consist of two stiff, parallel clamps with fibers between them.  All the fibers
have the same elastic constant. However, the maximum force each fiber can sustain before it fails irreversibly is set by a 
threshold drawn for each fiber from a probability distribution. Due to irreversibility, detailed balance does not hold.  Therefore, 
fiber bundle models are non-equilibrium systems. We focus here on the dynamical description of the Equal Load Sharing model (ELS) which 
is the mean field (MF) limit of the fiber bundle models \cite{skh15}. This is the model we sketched earlier in this paragraph. The equal load sharing fiber bundle model may be described through a discrete recursion relation (recursion dynamics) \cite{bpc03}. It was shown that 
ELS undergoes a phase transition.

The aim of this work is to derive a mesoscopic equation encapsulating the ELS critical dynamics. We show the close formal connection, in the limit of vanishing external field, between ELS critical behavior and an APT process, the Compact Directed Percolation (CDP) model. Next, we show that the ELS mesoscopic equation can be derived as a purely relaxational model depending on a Hamiltonian describing the ELS stationary behavior. Then, based on symmetry argument, we highlight the origin of the analogy between FBM and spinodal nucleation \cite{zrsv97,ku83}.

We will in the next section review the critical properties of the fiber bundle model using recursive dynamics. In section \ref{meso_els}
we introduce a mesoscopic description of ELS. We numericaly compare ELS and the mesoscopic equation, section \ref{data_collapse}. We proceed in section \ref{sec:dp} to show the phenomenological and formal similarities between ELS and CDP in an external field. Section \ref{sec_lang} demonstrates how the ELS fiber bundle model may be described through an overdamped Langevin equation. Then, in section \ref{sec:nuc} we explicitly link FBM and spinodal nucleation.  
In the last section, we summarize and discuss our work.

\section{Recursive dynamic of Equal Load Sharing Fiber Bundle Model} 
\label{sec:els}

The ELS fiber bundle model describes the breaking process of $N$ initially intact fibers subject to a homogeneous external 
field $\sigma$, the initial load per fiber. A fiber $j$ is characterized by a strength $\tau_j$ which is a threshold value sampled 
from a probability distribution $p$. We denote by $n(t)$ the number density (number of surviving fibers divided by $N$) of intact fibers at time $t$, $k(t)=1-n(t)$ the density of broken fibers (the damage) with $n(0)=1$.   
The dynamics of the system under load is defined as follows. At discrete time $t>0$, all fiber $j$ such that
\begin{equation}
\tau_j<\frac{\sigma}{n(t-1)}
\end{equation}
breaks irreversibly. Then, the number of intact fibers is updated and the process continue until the system reaches a stationary configuration. We can notice, that by definition the model is infinitely dimensional, $i.e.$ space plays no role. Local load redistribution introduces spatial effect in Fiber Bundle model and are studied for example in the local load sharing fiber bundle model 
where the nearest neighbors of the failed fibers absorb the load they were carrying at failure \cite{phc10,hhp15}.

The control parameter of ELS is $\sigma$. As we will see below, ELS exhibits critical behavior close to the critical point $\sigma=\sigma_c$. The exponents characterizing the system in vicinity of the critical point do not depend on the choice of the threshold probability, see \cite{phc10,hhp15,zrsv97}. In the following we will work with the uniform threshold distribution for simplicity.

Formally, the system dynamic is described by a recursive relation \cite{phc10,hhp15}. The density of broken fiber $k(t)$ is given by the threshold cumulative distribution $P(\tau)=\int_0^{\tau} p(\tau')d\tau'$,
\begin{equation}
k(t)= P\left(\frac{\sigma}{n(t)}\right)=\int_0^{\sigma/n(t)}p(x)dx\;.
\label{eq:els_k}
\end{equation}
Thus,
\begin{equation}
n(t+1)=1-P\left(\frac{\sigma}{n(t)}\right)=1-\frac{\sigma}{n(t)}
\label{eq:rec}
\end{equation}
since $P(\tau)=\tau$ for the uniform distribution on the unit interval. The breaking process occurs until the system reaches a stationary configuration
\begin{equation}
n_*=1-\frac{\sigma}{n_*}\;,
\end{equation}
with $n_*$ the stationary solution of equation (\ref{eq:rec}).
Therefore, the equation of state for the stable system is
\begin{equation}
n_*^2-n_*+\sigma=0\;.
\label{els_statio}
\end{equation}
Defining the system order parameter as $\eta=n_*-1/2$ \cite{phc10,hhp15}, we observe that
\begin{equation}
\eta\sim \left(\sigma_c-\sigma\right)^{\beta}\;,
\label{eq:statio_els}
\end{equation}
with $\sigma_c=1/4$ and $\beta=1/2$ as the order parameter exponent. Thus, to keep $\eta$ real, we study the system for load $\sigma\leq \sigma_c$.

At the critical point, i.e.\ at $\sigma=\sigma_c$ and for $t\rightarrow\infty$, given equation (\ref{eq:rec}), we have
\begin{equation}
\eta\sim t^{-\alpha}\;,
\label{eq:time_els}
\end{equation}
with $\alpha=1$.  This characterizes the critical slowing down of the fiber bundle model. 

Other standard universal exponents are found in the same way. The susceptibility is
\begin{equation}
\chi=\left|\frac{\partial \eta_*}{\partial \sigma}\right|\sim\left(\sigma_c-\sigma\right)^{\beta-1}=\left(\sigma_c-\sigma\right)^{-\gamma}\;,
\label{eq:suc}
\end{equation}
with $\gamma=1/2$ the susceptibility exponent. Note that, since the model is governed by only one physical parameter which is the external
load $\sigma$, the susceptibility exponent depends directly on the order parameter. The relaxation time $\xi_{\parallel}$ toward a stationary solution follows \cite{phc10}
\begin{equation}
\xi_{\parallel}\sim \left(\sigma_c-\sigma\right)^{-\nu_{\parallel}}\;,
\end{equation}
with $\nu_{\parallel}=1/2$ being the time correlation length exponent. We summarize the ELS universal exponents in Table \ref{tab:expo}.

\begin{table}[ht]
\center
\begin{tabular}{c|c}
&ELS\\
\hline\hline
$\beta$ & 1/2 \\
$\gamma$ & 1/2 \\
$\nu_{\parallel}$ & 1/2 \\
$\alpha$& 1 \\
\end{tabular}
\label{tab:dp_ci}
\caption{Mean field exponents of Fiber Bundle Model (ELS) using standard notation. 
\label{tab:expo}}
\end{table}

The continuous limit of equation (\ref{eq:rec}) can be readily obtained and is \cite{phc10}
\begin{equation}
\frac{\partial n}{\partial t}=-\frac{n^2-n+\sigma}{n}\;.
\label{eq:conti_els}
\end{equation}
The presence of the density in the denominator of this equation makes it hardly amenable for a standard field theory treatment. In the following, we introduce an alternative mesoscopic equation.

\section{Mesoscopic ELS equation}
\label{meso_els}

We show how to simplify, keeping the same critical behavior, the ELS continuous equation (\ref{eq:conti_els}). We have, introducing the order parameter $\eta=n-1/2$ in the last equation
\begin{equation}
\frac{\partial \eta}{\partial t}=-\frac{\eta^2-J}{\eta+1/2}\;,
\label{eq:conti_els_order}
\end{equation}
$J=\sigma_c-\sigma=1/4-\sigma\;.$
In the double limit $\eta<<1/2$ and $J\rightarrow 0$, we can write
\begin{equation}
\frac{\partial \eta}{\partial t}\approx2(-\eta^2+J)\;.
\label{eq:conti_els_order_approx}
\end{equation}
We observe, in the limit of $t\rightarrow\infty$, that we recover exactly stationary behavior of the ELS model. We further simplify the equation by absorbing the factor $2$ in the time parameter,
\begin{equation}
\frac{\partial \eta}{\partial t}=-\eta^2+J\;.
\label{eq:ci}
\end{equation}
This generalized equation encapsulates the mesoscopic behavior of the ELS model.

Indeed, for example, the order parameter exponent is given by
\begin{equation}
\eta_*\sim J^{\beta}\;,
\end{equation}
with $\eta_*$ the stationary solution and $\beta=1/2$. At criticality, $i.e.$ for $J=0$, we have
\begin{equation}
\eta_c(t)\sim t^{-\alpha}\;,
\end{equation}
$\alpha=1$. The susceptibility and time correlation length can also easily be computed, see for example \cite{o08}. It 
appears that equation (\ref{eq:ci}) reproduces the universal exponents of the ELS model. The scaling forms of ELS and the mesoscopic equation (\ref{eq:ci}) are studied in the next section.

Equation (\ref{eq:ci}) coincide (among other microscopic interpretations) with the mean field rate behavior of a coagulation with input (CI) reaction diffusion process, where $\eta$ is the particle density. More precisely, CI describes particles $A$ diffusing on a lattice that coagulate when they meet ($A+A\longrightarrow A$) with a source term ($\emptyset\longrightarrow A$) acting at rate $J$. This system is characterized by an upper critical dimension $d_c^u=2$. It was extensively studied by Droz and coworkers, see \cite{ds93,rd97}. 
The mean field rate Equation (\ref{eq:ci}) holds for CI above $d=2$. 

This formal similarity between CI and FBM invites us 
to formulate the latter using reaction-diffusion or epidemic propagation phenomenon terminology, 
see section \ref{sec:dp}. 

\section{Data collapse} 
\label{data_collapse}

In the last section, we showed that the mesoscopic description is obtained as the limit $t\rightarrow\infty$ and/or $J\rightarrow 0$ of the ELS dynamics.

By definition two systems belong to the same universality class if they have the same critical exponents and, near the critical point, if their scaling functions are identical. In this aim, we numerically solve and compare equations (\ref{eq:conti_els}) and (\ref{eq:ci}). Remark: meanwhile the exact solution of equation (\ref{eq:ci}) is easy to obtain, a numerical approach is needed for the ELS one. Hence, we carry out the data collapse for different solutions close to the critical point. 

We note $\eta_{\text{els}}\left(t,J\right) = n_{\text{els}}\left(t,J\right) -1/2$ is the shifted solution of (\ref{eq:conti_els}) with $J=1/4-\sigma$ and $\eta_{\text{meso}}\left(t,J\right)$ the solution of (\ref{eq:ci}). We take the initial configurations $n_{\text{els}}=1=\eta_{\text{meso}}$. Both solutions, near criticality, behave as
\begin{equation}
\eta=\left(a_t t\right)^{-\alpha}\lambda R\left(J\left(a_tt\right)^{1/{\nu_{\parallel}}}\right)
\label{eq:scal}
\end{equation}
with $R$ and $a_t$ the corresponding ELS or mesoscopic, scaling functions and non-universal metric factors. Normalizing $R$ by $R(0)=1$, we can find the metric factors as the amplitude of the power law at the critical point $J=0$. We obtain $a_t^{\text{ELS}}=1/2$ and $a_t^{\text{meso}}=1$ (see equation (\ref{eq:conti_els_order_approx})).

\begin{figure}[hbt]
\includegraphics[scale=0.45]{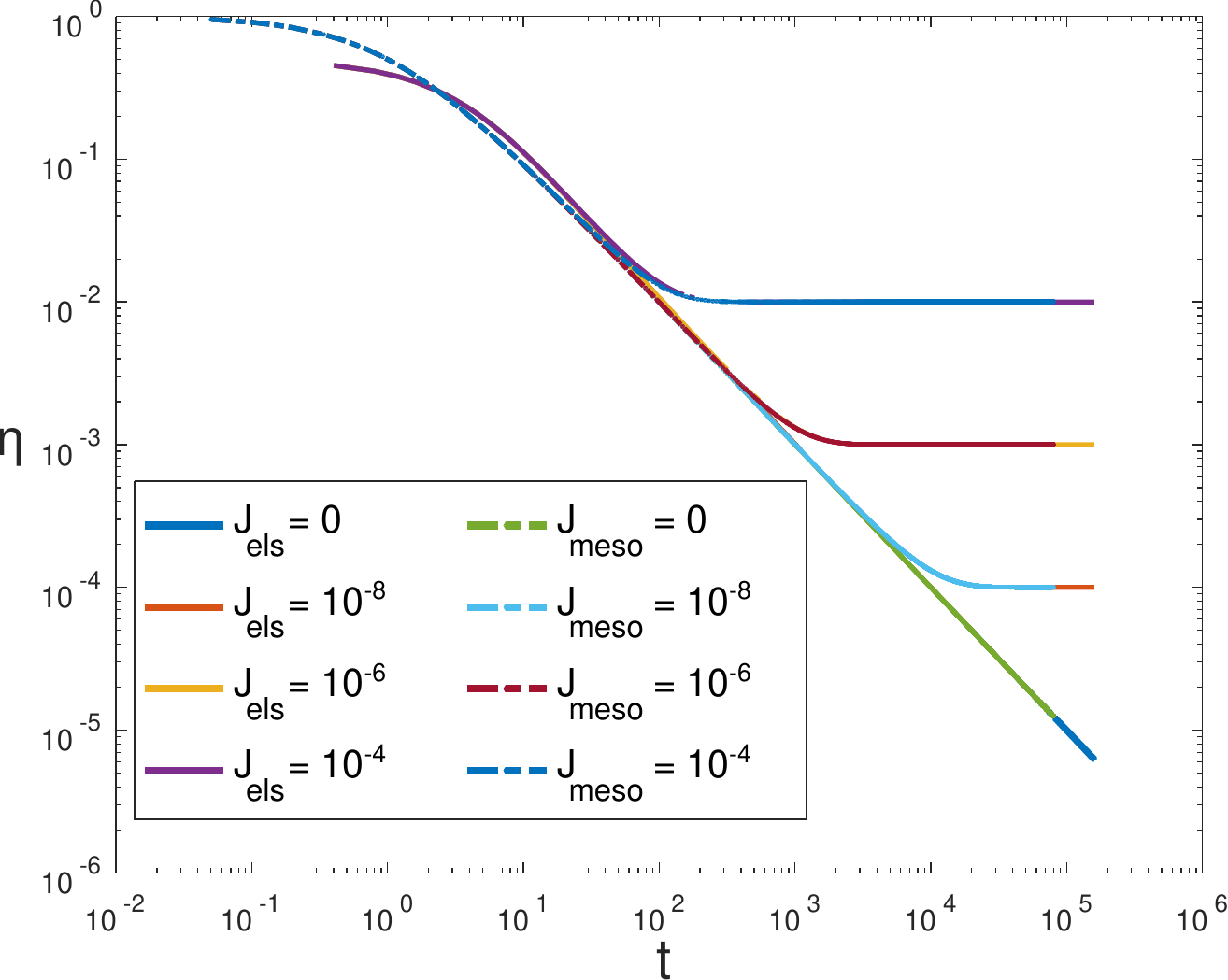}
\caption{Solutions of the ELS continuous equation (\ref{eq:conti_els}) and mesoscopic equation (\ref{eq:ci}) with the time rescaled by $t\longrightarrow t/a_t$.}
\label{fig:sol}
\end{figure}

\begin{figure}[hbt]
\includegraphics[scale=0.45]{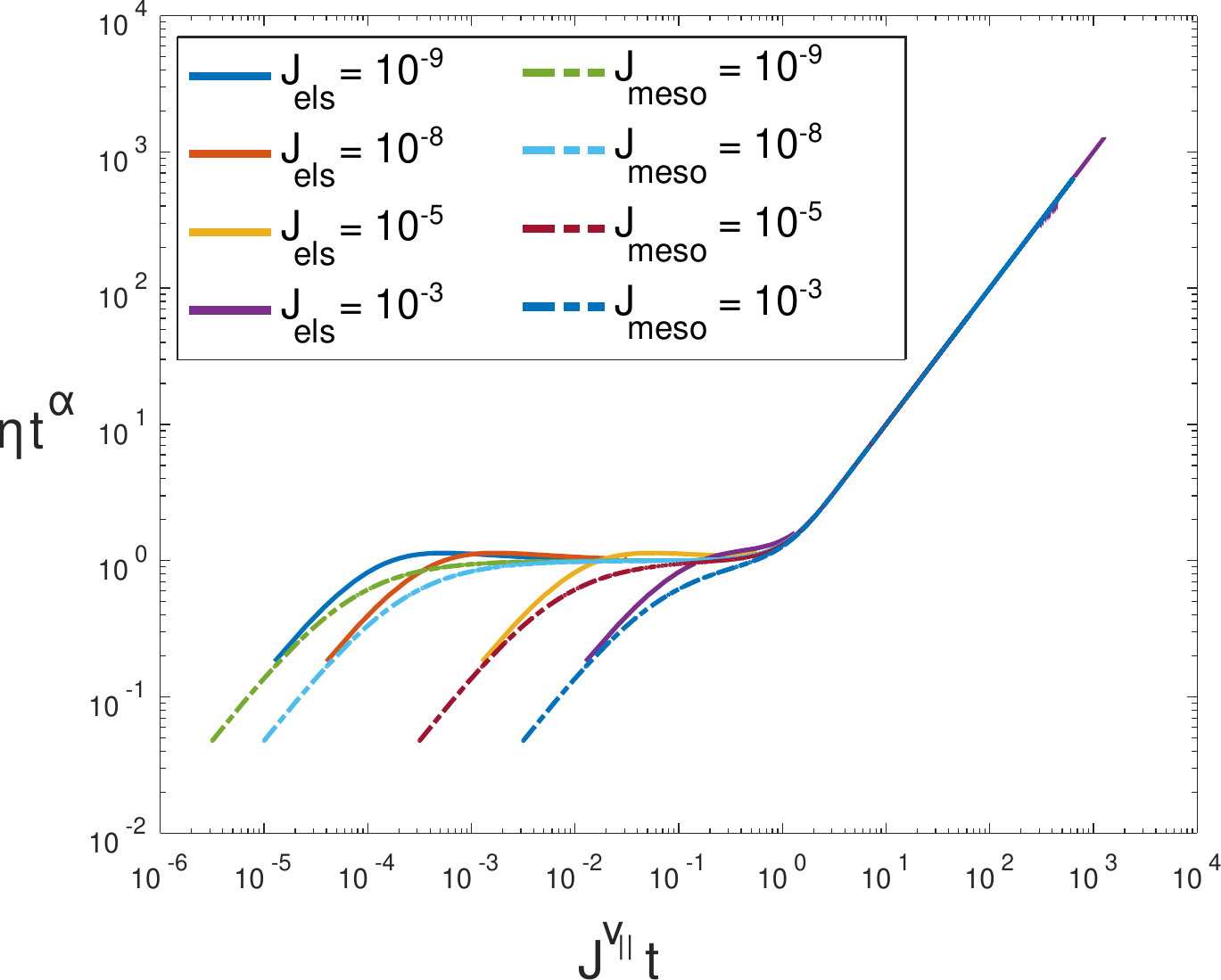}
\caption{Data collapse according to the scaling form (\ref{eq:scal}) with $\alpha=1$ and $\nu_{\parallel}=1/2$. The time is rescaled by $t\longrightarrow t/a_t$.}
\label{fig:data_coll}
\end{figure}

To compare $R^{\text{ELS}}$ and $R^{\text{meso}}$ we rescale, for the ELS and the mesoscopic equation, the time by the corresponding metric factors $i.e,$ $t\longrightarrow t/a_t$. The solutions are presented in Figure \ref{fig:sol}. 
The data collapse is shown in Figure \ref{fig:data_coll}. It provides a convincing hint that the the mesoscopic equation and the ELS model belong to the same universality class.\footnote{The exact solution of 
(\ref{eq:ci}) for a non-zero initial condition $\eta(0)>0$ gives a small imaginary contribution. Note that we do not observe it numerically, using a standard ODE solver. We neglect it as 
it does not contribute to the singular behavior of the transition. The imaginary part decreases exponentially with time and, thus, the stationary solution is real since we take $J\geq0$. Moreover, the critical solution $J=0$ is also real ($\eta_{\text{meso}}(t,J=0)=1/(1+t)$). Finally, since we do not have an exact solution for the ELS equation (\ref{eq:conti_els}) nothing guarantees that the solutions are also real. However, as for the mesoscopic equation, we do not observe any imaginary parts through numerical solutions.}
 
The differences observed between the two models are consistent with the considered approximations to derive the mesoscopic equation (\ref{eq:ci}). Remark: The term ``mesoscopic" is justified since it 
captures the critical behavior of the  ELS model for large time and/or small control parameter $J$.

\section{Fiber Bundle Model as an Absorbing Phase Transition}
\label{sec:dp}

In absence of conservation laws or specific symmetries, it is expected that non-equilibrium phase transitions from an active to a unique absorbing state are described by the directed percolation universality class \cite{j81,g82,t14}. Here we show that the mesoscopic ELS behavior, in zero external field, is characterized by a particular symmetry, namely the particle-hole symmetry.

The large scale behavior of ELS is encompassed by the order parameter mesoscopic equation (\ref{eq:ci}). Rewriting it in term of the density of intact fibers $n$, $i.e.$ $\eta=n-n_c$ ($n_c=1/2$), we have
\begin{equation}
\frac{\partial n}{\partial t}=\lambda n\left(1-n\right)-\sigma\;,
\label{eq:cdp}
\end{equation}
with $\lambda=1$. In zero external field, $\sigma=0$, the last equation obeys the particle-hole symmetry, $i.e.$
\begin{align*}
n&\longleftrightarrow 1-n\;,\\
\lambda &\longleftrightarrow -\lambda\;.
\label{eq:phs}
\end{align*}
The invariance under this transformation is characterising the Compact Directed Percolation (CDP) universality class \citep{hhlp08,j05} (also named compact domain growth). The CDP upper critical dimension is $d_c=2$. For $d>2$, the process is exactly described by equation (\ref{eq:cdp}) with $\sigma=0$. In a CDP process, the dynamics occurs only at boundaries between clusters of active and inactive sites. In other words, using the FBM terminology, the dynamics take place at the domain walls between clusters of intact and broken fibers. Moreover, fibers cannot break (or be created, due to irreversibility) spontaneously, they break only under load redistribution which occurs at the boundaries between clusters of broken and intact fibers. 

In this work we focus on ELS, and thus, we do not have access to spatial features of the FBM. However, Local Load Sharing (LLS) model \cite{skh15}, an FBM for which load redistribution acts locally and whose MF limit is ELS, is precisely characterized by a dynamics that takes place at the interfaces between clusters of intact and broken fibers. Thus, at a phenomenological level, the attempt to relate ELS with CDP is founded. However, in FBM, unlike in CDP, a broken fiber cannot recover. We expect that it does not impact, in average, on large scale behavior of the order parameter.

Here, we clarify the role played by the external field $\sigma$. The ELS model is driven by $\sigma$ which breaks the particle-hole symmetry. The external field $\sigma$ initiates the primary holes, the seeds, around which load redistribution take place. Indeed, equation (\ref{eq:rec}) at step $t=0$ gives $n(1)=1-\sigma$ considering, initially, all fiber intact, $n(0)=1\;.$ Then, the dynamics occurs first around the seeds and then around the germinate clusters of broken fibers until the system reach a stable configuration given $\sigma$.

In this Section, we aimed to link the FBM with CDP a well-known process that undergoes an absorbing phase transition. We took advantage of the symmetry of the mesoscopic equation in zero external field to compare formally and phenomenologically ELS with CDP. However, since we work at the MF level we cannot elaborate an unique field theory describing FBM. A mapping of LLS model to an APT process is needed.

We note that the external field $\sigma$ destroys the absorbing phase and reduces it to a point. Considering $\eta=n-n_c$ as the order parameter, the absorbing point is located at $\sigma=\sigma_c=1/4$. In zero field, CDP has two absorbing states, the empty lattice and the fully occupied lattice reflecting the emerging $\mathbb{Z}_2$ symmetry.

\section{Equal Load Sharing as an overdamped Langevin Equation} 
\label{sec_lang}

The ELS fiber bundle model in its dynamical formulation exhibits a dynamical phase transition. It provides, in the vicinity of the 
critical point, a natural process that ensures a time scale separation between the kinetic of the order parameter and the renaming 
physical quantities \cite{t14}. As we will see, the ELS model can be described through an overdamped Langevin equation of the form
\begin{equation}
\frac{\partial \eta(t)}{\partial t}=-\frac{\delta H\left[\eta\right]}{\delta \eta}+\zeta(t)\;,
\label{eq:relax_dyn}
\end{equation}
with $H$ the Hamiltonian of the system and $\zeta(t)$ a noise term. Since, by essence, ELS is mean field we set $\zeta(t)=0$.

\subsubsection*{Derivation of H}
\label{sub_derh}

The ELS and other versions of the fiber bundle model are mainly studied in their quasistatic limit. In this picture, fibers are 
broken one by one. The quasistatic limit corresponds to slowly stretching the fibers. Therefore, it is convenient to introduce the fiber elongation $x=\sigma/n$ since it is assumed that fibers behave as Hookean springs.

The ELS energy contents, defined through the work done on the fiber bundle under increasing load, was recently 
studied by Pradhan et al.\ \cite{phr18}. The total energy content at damage $k=1-n$ and elongation $x$ is
\begin{equation}
H\left[x,k\right]=\theta\left(x^2(1-k)+\int_0^k d\delta \left(P^{-1}(\delta)\right)^2\right)\;,
\label{eq:els_hamilto}
\end{equation}
with $\theta=N\kappa/2$, $\kappa$ the Hookean constant, $P^{-1}(\delta)$ the inverse function of the threshold cumulative distribution. The first term of equation (\ref{eq:els_hamilto}) is the Hookean energy, the second one is the energy dissipated through 
fiber failures and hence responsible for the formation of holes, that is clusters of broken fibers.

The equation of state of the system is
\begin{equation}
0=\frac{\delta H\left[x,k\right]}{\delta k}=\theta\left(-x^2+\left(P^{-1}(k)\right)^2\right)\;,
\end{equation}
and thus, since $x$ and $P^{-1}(k)$ are positive quantities
\begin{equation}
x=P^{-1}(k)=k\;,
\label{eq:state}
\end{equation}
assuming a uniform threshold distribution on the unit interval. By definition of the elongation we have
\begin{equation}
x=\frac{\sigma}{n_*}=k\;.
\end{equation}
Hence, using $k=1-n_*$, we obtain again equation (\ref{els_statio}), the equation of state of ELS 
\begin{equation}
n_*^2-n_*+\sigma=0\;.
\label{eq:uni1}
\end{equation}
We are interested in catching the stationary critical behavior of the system at the Hamiltonian level in an easier to handle 
expression than equation (\ref{eq:els_hamilto}). Since the system equation of state is given by equation (\ref{eq:uni1}), we can write
\begin{equation}
\frac{\delta H\left[n_*,\sigma\right]}{\delta n_*}=n_*^2-n_*+\sigma=0\;.
\label{eq:uni}
\end{equation}
Integrating this last expression, we observe that
\begin{equation}
H\left[\eta,J\right]=\eta^3/3-J\eta\;,
\label{eq:hamilto_ci}
\end{equation}
with $\eta=n_*-1/2$ and $J=\sigma_c-\sigma$.
We note that this Hamiltonian was previously studied in the context of spinodal nucleation, see section \ref{sec:nuc}.

\subsubsection*{ELS overdamped Langevin equation}
\label{sub_overd}

The ELS critical slowing down, equation (\ref{eq:time_els}), provides us with a natural time--length scale separation between the order
parameter kinetics and the other physical quantities. These quantities appear as surrounding noise from the order parameter's 
point of view. Therefore, the critical dynamic of ELS may be described by equation (\ref{eq:relax_dyn}). By inserting the Hamiltonian 
(\ref{eq:hamilto_ci}) into this equation, we find
\begin{equation}
\frac{\partial \eta}{\partial t}=-\eta^2+J\;,
\end{equation}
which is the previously introduced mesoscopic equation (\ref{eq:ci}). 

This approach is directly inspired by the time-dependent Ginzburg-Landau equation which is also named model A dynamics \cite{t14}
describing, for example, the Glauber model. Glauber dynamics is a minimal kinetic extension of the Ising model for which detailed balance ensures that the system relaxes toward the Ising canonical equilibrium probability distribution without conservation of the order parameter. Remark: in ELS the order parameter is also not a conserved quantity since fibers break to eventually reach a stable configuration, this justify equation (\ref{eq:relax_dyn}).

However, unlike the model A, ELS is a genuine non-equilibrium process. Thus, there is no detailed balance relation that can be used to derive a noise term for equation (\ref{eq:relax_dyn}).
\section{Nucleation Field Theory and ELS}
\label{sec:nuc}

In \cite{zrsv97}, analogy between ELS and spinodal nucleation \cite{ku83} has been observed. By studying the mesoscopic behavior of ELS, we show explicitly the underlying reason which rely on particle-hole invariance of the mesoscopic ELS equation. 

The Hamiltonian (\ref{eq:hamilto_ci}) is formally equal to the field-theoretic description of spinodal nucleation \cite{ku83}. More explicitly, the Landau-Ginzburg-Wilson Hamiltonian in the mean field approximation is the free energy
\begin{equation}
F\left[\psi\right]=b\psi^2+c\psi^4-h\psi
\label{eq:lg}
\end{equation}
with $b=a( T-T_c)$ the distance from the critical temperature $T_c$, $a>0$ and $c>0$ two constants.
In zero external fielf $h=0$ and for $T<T_c$ ($i.e.$ $b<0$), $F$ exhibits the two characteristic symmetric wells. Increasing $h>0$, the well located in negative $\psi$ values, become shallower and eventually disappears at $h_s$ and $\psi_s=-\left(|b|/6c\right)^{1/2}$, the spinodal. Then, close to the spinodal, introducing the field $\phi=\psi-\psi_s$, neglecting the irrelevant $\phi^4$ term \cite{ku83}, we have
\begin{equation}
F_s\left[\phi\right]=\epsilon\phi-\alpha\phi^3
\label{eq:spinodal}
\end{equation}
with $\epsilon\propto h_s-h$ and $\alpha$ a positive constant depending on $b$ and $c$. The mean-field spinodal theory (\ref{eq:spinodal}) is formally equivalent to equation (\ref{eq:hamilto_ci}). 

In section \ref{sec:dp}, we showed that the ELS at the mesoscopic level in zero external field is invariant under the particle-hole symmetry. It allows us to employ the $\mathbb{Z}_2$ invariant expression (\ref{eq:lg}) to describe zero field ELS. Indeed, introducing the lattice-gas mapping $\psi=2n-1$, we can readily show that the spin sign symmetry is equivalent to the particle-hole symmetry for lattice-gas
\begin{equation}
\psi\longleftrightarrow-\psi\;\;\;\Longleftrightarrow\;\;\; n\longleftrightarrow 1-n\;.
\end{equation}
The lattice-gas variable is suitable to describe ELS since we focus on particle density instead of magnetisation.

To conclude, we argued and showed, relying on the particle-hole symmetry, why ELS model can be viewed as a (genuine non-equilibrium) realization of spinodal nucleation.

\section{Discussion}
\label{sec_finale}

In this work, we derive a mesoscopic equation describing ELS critical behavior. The mesoscopic equation is invariant under particle-hole symmetry in zero external field. It enables us to formally link ELS with Compact Directed Percolation model which describe processes characterized by dynamics occurring at intact broken clusters' boundaries (which is precisely how FBM dynamic evolves). Then, we describe the mesoscopic behavior of ELS as an overdamped Langevin equation. The outcome is the derivation of an Hamiltonian representing ELS stationary behavior. Due to the particle-hole symmetry, we explicit the formal ELS/spinodal nucleation equivalence.

One of the main outcomes of this work is to link the ELS fiber bundle model, and more generally damage models, to the powerful formalism developed to study non-equilibrium phase transitions. It opens the way for a field theoretical treatment of such models. In this work we focus on ELS, due to its mean field nature it is not possible to define an unique field theory for FBM. Hence, in the future, we will concentrate our work on space dependent fiber bundle models such as the Local Load Sharing model.

\begin{acknowledgments}
The authors thank Jonas T.\ Kjellstadli for interesting discussions.  This work was partly supported by the
Research Council of Norway through its Centers of Excellence funding
scheme, project number 262644. M.\ H.\ thanks the Swiss National Science Foundation for an early postdoc mobility grant, 
number 171982.
\end{acknowledgments}

\end{document}